\documentclass[twocolumn,prb]{revtex4}
\usepackage{amsmath}
\usepackage{amsfonts}
\usepackage{amssymb}
\usepackage{graphics}
\usepackage{float}
\usepackage{graphicx}
\usepackage{epstopdf}
\usepackage[compact]{titlesec}
\usepackage{natbib}
\usepackage{threeparttable}

\begin{document}
\title{Tunable chirality and circular dichroism of a topological insulator with $ C_{2v} $ symmetry as a function of Rashba and Dresselhaus parameters}
\author{Parijat Sengupta}
\email{parijats@bu.edu}
\author{Enrico Bellotti}
\affiliation{Photonics Center \\
8 St. Mary's Street, Boston University \\
Boston, MA, 02215.}

\begin{abstract}
Polarization-sensitive devices rely on meta-materials to exhibit varying degrees of absorption of light of a given handedness. The chiral surface states of a topological insulator(TI) selectively absorb right and left circularly polarized light in the vicinity of the Dirac cone reaching its maximum of unity at the $ \Gamma $ point. In this letter, we demonstrate that a band gap open TI with C$_{2v}$ symmetry which is represented through a combination of Rashba and Dresselhaus Hamiltonians alters the preferential absorption of left and right circularly polarized light allowing a smooth variation of the circular dichroism(CD). This variation in CD, reflected in a range of positive and negative values is shown to be a function of the Rashba and Dresselhaus coupling parameters. Additionally, we draw a parallel between the varying CD and the emerging field of valley-electronics in transition metal dichalcogenides. 
\end{abstract}
\maketitle

Optical chirality is manifest through circular dichroism~\cite{cao2012valley,yao2008valley}(CD) which relates to a differential absorption of left and right circularly polarized light and is a key component of chiral designer meta-materials used in polarization sensitive imaging devices and display technologies. Topological insulators(TI) with their helical spin structure offer an active control over chiral handedness observed through a varying degree of polarization-dependent absorption. In this letter, we show that in a band gap open topological insulator with C$_{2v}$ symmetry which serves as a model meta-material, circular dichroism can be smoothly varied without any microscopic reconfiguration of the surface. We demonstrate that circular  dichroism measured by the degree of circular polarization can assume both positive and negative values. This finding suggests that left and right circularly polarized light can be selectively absorbed in the vicinity of the Dirac point by a suitable adjustment of the Rashba and Dresselhaus parameters. 

To begin, we first introduce the model Hamiltonian for the band gap open surface states of a thin TI slab with C$_{2v}$ symmetry. The surface states of the TI slab, for the chosen symmetry, are described by a combination of the Rashba and Dresselhaus spin-orbit Hamiltonians.~\cite{winkler2003spin,fabian2007semiconductor} Using the corresponding wave functions, we evaluate circular dichroism around the Dirac point (at $ \Gamma $). The band gap of the surface states is considered to be a phenomenological parameter arising due to a coupling of the top and bottom surfaces of the thin TI film. The tuning of circular dichroism is taken up next followed by a brief discussion on connection to the emerging field of valley-electronics~\cite{wang2013valley,konabe2014valley} in transition metal dichalcogenides(TMDCs). 

The Hamiltonian that describes the surface states of such a TI slab is
\begin{equation}
H_{TI} = \alpha_{R}\left(\sigma_{y}k_{x} - \sigma_{x}k_{y}\right) + \alpha_{D}\left(\sigma_{x}k_{x} - \sigma_{y}k_{y}\right) + \Delta\sigma_{z}, 
\label{ham1} 
\end{equation}
where $ \alpha_{R} $ and $ \alpha_{D} $ are the Rashba and Dresselhaus parameters, respectively.
The two-component wave functions for the Hamiltonian $ H_{TI} $ are
\begin{subequations}
\begin{equation}
\Psi_{\pm} = \dfrac{1}{\sqrt{2}}\begin{pmatrix}
\lambda_{\pm}exp\left(-i\theta\right) \\
\lambda_{\mp}  
\end{pmatrix},
\label{wf1}
\end{equation}
where $ \lambda_{\pm} $ are defined as
\begin{equation}
\lambda_{\pm} = \sqrt{1 \pm \dfrac{\Delta}{\sqrt{\Delta^{2}+\vert \beta \vert^{2}}}}.
\label{wf2}
\end{equation}
The polar angle $ \theta $ and $ \beta $ in Eq.~\ref{wf1} and Eq.~\ref{wf2}, respectively are 
\begin{equation}
\theta = tan^{-1}\dfrac{k_{x}+ \kappa\,k_{y}}{k_{y}+ \kappa\,k_{x}} =  tan^{-1}\dfrac{cos\phi+ \kappa\,sin\phi}{sin\phi+ \kappa\,cos\phi},
\label{polaz}
\end{equation}
and
\begin{equation}
\vert \beta \vert^{2} =  k^{2}\left(\alpha_{R}^{2} + \alpha_{D}^{2} + 2\alpha_{R}\,\alpha_{D}\,sin\,2\phi\right),
\label{betal}
\end{equation}
where $ \kappa = \alpha_{D}/\alpha_{R} $, the ratio of the Dresselhaus and Rashba coupling coefficients. Also, $ k_{x} = k\,cos\phi $ and $ k_{y} = k\,sin\phi $.
\end{subequations}

\noindent The circular dichroism that we seek to compute in this paper is defined as~\cite{ezawa2012spin}
\begin{subequations}
\begin{equation}
\eta\left(k\right) = \dfrac{\vert \mathcal{P}_{+}\left(k\right)\vert^{2}-\vert \mathcal{P}_{-}\left(k\right)\vert^{2}}{\vert \mathcal{P}_{+}\left(k\right)\vert^{2}+\vert \mathcal{P}_{-}\left(k\right)\vert^{2}},
\label{cd1}
\end{equation}
where $ \mathcal{P}_{\pm}\left(k\right) $ are described in terms of the inter band matrix elements as $ \mathcal{P}_{\pm}\left(k\right) = \mathcal{M}_{cv}^{x} \pm i\mathcal{M}_{cv}^{y} $. The inter band matrix elements are expressed in the usual way, for instance, $ \mathcal{M}_{cv}^{x} $ is
\begin{equation}
\mathcal{M}_{cv}^{x} = \langle\Psi_{+}\vert \hat{v}_{x}\vert \Psi_{-}\rangle.
\label{interband}
\end{equation}
\end{subequations} 
We therefore write out the velocity components corresponding to the Hamiltonian of Eq.~\ref{ham1}. The velocity component directed along \textit{x}-axis is
\begin{subequations}
\begin{flalign}
v_{x} &= \dfrac{1}{\hbar}\dfrac{\partial\,H}{\partial\,x} 
= \dfrac{1}{\hbar}\begin{pmatrix}
0 & \alpha_{D} + i\alpha_{R}  \\
\alpha_{D} - i\alpha_{R} & 0
\end{pmatrix}.
\label{velx}
\end{flalign}
Similarly the velocity component along \textit{y}-axis is
\begin{flalign}
v_{y} &= \dfrac{1}{\hbar}\dfrac{\partial\,H}{\partial\,x} = \dfrac{1}{\hbar}\begin{pmatrix}
0 & \alpha_{R} + i\alpha_{D} \\
\alpha_{R} - i\alpha_{D} & 0
\end{pmatrix}.
\label{vely}
\end{flalign}
\end{subequations}

\noindent The inter band matrix elements $ \mathcal{M}_{cv}^{x} $ and $ \mathcal{M}_{cv}^{y} $ can be easily expressed using Eqs.~\ref{interband},~\ref{velx}, and ~\ref{vely} yielding
\begin{subequations}
\begin{align}
\mathcal{M}_{cv}^{x} &= \dfrac{1}{2\,\hbar}\begin{pmatrix}
\lambda_{+}\,exp\left(i\theta\right) & \lambda_{-} 
\end{pmatrix}v_{x}\begin{pmatrix}
\lambda_{-}\,exp\left(-i\theta\right)\\
-\lambda_{+}
\end{pmatrix} \notag \\
&= \dfrac{1}{\hbar}\left[-i\,\left(\alpha_{D}\,sin\theta  + \alpha_{R}\,cos\theta\right) \right. \notag \\
&\left. -\dfrac{\Delta}{\sqrt{\Delta^{2} + \vert \beta \vert^{2}}}\left(\alpha_{D}\,cos\theta -\alpha_{R}\,sin\theta\right)\right] ,
\label{mx1} 
\end{align}
and
\begin{align}
\mathcal{M}_{cv}^{y} &= \dfrac{1}{2\,\hbar}\begin{pmatrix}
\lambda_{+}\,exp\left(i\theta\right) & \lambda_{-} 
\end{pmatrix}v_{y}\begin{pmatrix}
\lambda_{-}\,exp\left(-i\theta\right)\\
-\lambda_{+}
\end{pmatrix} \notag \\
&= \dfrac{1}{\hbar}\left[-i\,\left(\alpha_{R}\,sin\theta  + \alpha_{D}\,cos\theta\right) \right. \notag \\
&\left. - \dfrac{\Delta}{\sqrt{\Delta^{2} + \vert \beta \vert^{2}}}\left(\alpha_{R}\,cos\theta -\alpha_{D}\,sin\theta\right)\right].
\label{my1}
\end{align}
\end{subequations}

We now derive analytic expression (after some lengthy and tedious but simple algebra) for circular dichroism which is measured by computing the degree of circular polarization (Eq.~\ref{cd1}). We start with Eqs.~\ref{mx1} and ~\ref{my1} to write
\begin{subequations} 
\begin{flalign}
\vert \mathcal{P}_{+}\left(k\right)\vert^{2} &= \vert \mathcal{M}_{cv}^{x} + i\mathcal{M}_{cv}^{y}\vert^{2} \notag \\
&=\dfrac{1}{\hbar^{2}}\left[ \left(1 + \gamma\right)^{2}\alpha_{R}^{2} +  \left(1 - \gamma\right)^{2}\alpha_{D}^{2} \right. \notag \\
&\left. + 2\left(1 - \gamma^{2}\right)\alpha_{R}\,\alpha_{D}\,sin\,2\theta\right].
\label{pplus}
\end{flalign}.
Similarly, $ \vert \mathcal{P}_{-}\left(k\right)\vert^{2} $ is
\begin{flalign}
\vert \mathcal{P}_{-}\left(k\right)\vert^{2} &= \vert \mathcal{M}_{cv}^{x} - i\mathcal{M}_{cv}^{y}\vert^{2} \notag \\
&=\dfrac{1}{\hbar^{2}}\left[ \left(1 + \gamma\right)^{2}\alpha_{D}^{2} +  \left(1 - \gamma\right)^{2}\alpha_{R}^{2} \right. \notag \\
&\left. - 2\left(1 - \gamma^{2}\right)\alpha_{R}\,\alpha_{D}\,sin\,2\theta\right],
\label{pminus}
\end{flalign}
where $ \gamma = \dfrac{\Delta}{\sqrt{\Delta^{2} + \vert \beta \vert^{2}}} $.

\noindent The degree of circular polarization is therefore straightforward to obtain by inserting expressions for $ \vert \mathcal{P}_{\pm}\left(k\right)\vert^{2} $ in Eq.~\ref{cd1} and integrating out the angular dependence to give
\begin{align}
\eta\left(k\right) &= \dfrac{1}{\pi}\int_{0}^{2\pi}\dfrac{\gamma\left(\alpha_{R}^{2} - \alpha_{D}^{2}\right)+ \left(1 - \gamma^{2}\right)\alpha_{R}\,\alpha_{D}\,h\left(\phi\right)}{\left(1 + \gamma^{2}\right)\left(\alpha_{R}^{2} + \alpha_{D}^{2}\right)}d\,\phi. \notag \\
\label{etaint}
\end{align}
This can be more conveniently expressed as
\begin{equation}
\eta\left(k\right) = \dfrac{1}{\pi}\int_{0}^{2\pi}\dfrac{\gamma\left(1 - \kappa^{2}\right)+ \left(1 - \gamma^{2}\right)\kappa\,h\left(\phi\right)}{\left(1 + \gamma^{2}\right)\left(1 + \kappa^{2}\right)}d\,\phi,
\label{etafin}
\end{equation}
where $ \kappa = \alpha_{D}/\alpha_{R} $ and $ h\left(\phi\right)$ is $ sin\,2\theta $ expressed as a function of $ \phi $ using Eq.~\ref{polaz}. The integral in Eq.~\ref{etafin} can be numerically evaluated. Note that the factor of $ 1/\pi $ comes from averaging over all angles of incidence between 0 and $ 2\pi $ and $ \alpha_{R} $ is tacitly assumed to be non-zero. 
\end{subequations}

\begin{figure}[b]
\includegraphics[scale=0.9]{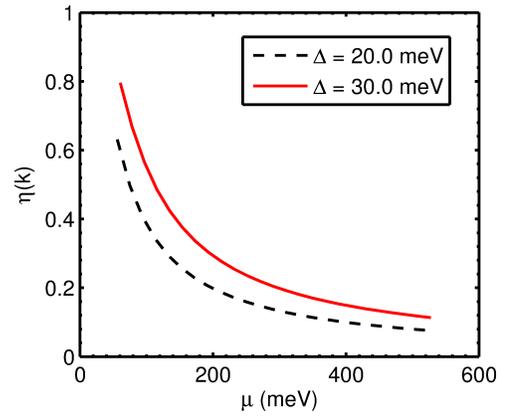} 
\caption{The degree of circular polarization $ \eta(k) $ plotted as a function of Fermi energy. The Fermi velocity v$_{f}$ is 0.8 $ \times $ 10$^{6}$ m/s. Circular dichroism measured through $ \eta(k) $ is more pronounced for the TI with a larger band gap since optical absorption increases.} 
\label{nodress}
\end{figure}

We first establish the circular dichroism dependence on the Fermi energy by ignoring the Dresselhauss component in Eq.~\ref{ham1} which reduces it to a band gap open linear Dirac Hamiltonian. The degree of circular polarization which is the indicator of circular dichroism, in this case, is plotted as a function of the Fermi energy $\left(\mu =  \sqrt{\hbar^{2}v_{f}^{2}k^{2} + \Delta^{2}} \right) $  by setting $ \alpha_{D} = 0 $ and  $ \alpha_{R} = \hbar\,v_{f} $ in the inter band matrix elements given in Eqs.~\ref{mx1} and ~\ref{my1}. The band gaps $ \Delta $ for two different cases was set to 20.0 $\mathrm{meV} $ and 30.0 $ \mathrm{meV} $. The noteworthy feature of Fig.~\ref{nodress} is that circular dichorism ($ \eta(k) $) goes to zero at \textit{k}-points away from the $ \Gamma $ point. This observation can be simply explained by taking recourse to the fact that the optical selection rule holds exactly at the $ \Gamma $ point where it is unity for a topological insulator but progresses to zero as $ k $ increases.

\begin{figure}
\includegraphics[scale=0.85]{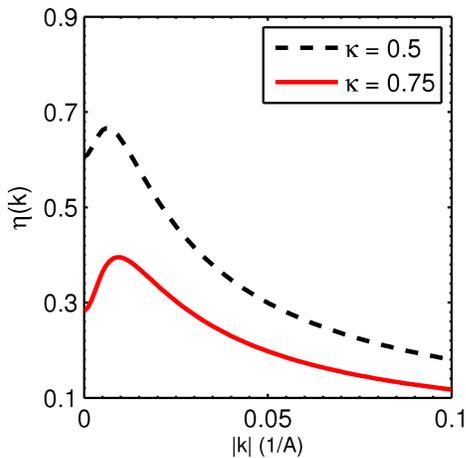} 
\caption{The degree of circular polarization $ \eta(k) $ is shown as a function of the $ k $ vector. The Fermi velocity v$_{f}$ is 5 $ \times $ 10$^{5}$ m/s. Circular dichroism is no longer unity at the $ \Gamma $ point indicating a composite spin texture arising from the Rashba and Dresselhaus spin Hamiltonians.} 
\label{rdallp}
\end{figure}

\begin{figure}[b]
\includegraphics[scale=0.85]{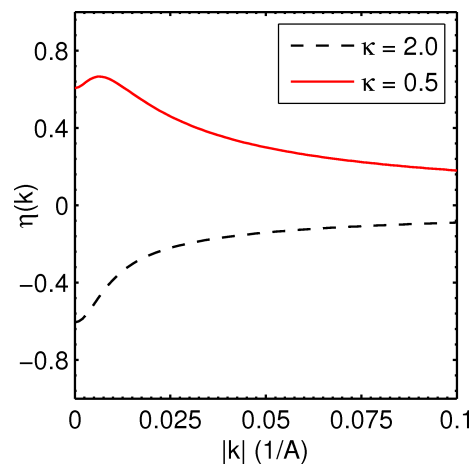} 
\caption{The dual nature of circular dichroism is clearly seen in this figure. Circular dichroism for the same energy state $\left(200.0\,\mathrm{meV}\right)$ at two different values of $ \kappa $ at $ k = 0 $ has a pair of values with reversed signs. } 
\label{rdpn}
\end{figure}

When both $ \alpha_{D} $ and $ \alpha_{R} $ are non-zero, the circular dichroism depends on their relative strength which is easy to understand if we recollect that for the chiral surface states of a topological insulator, light wave vector $ q $ couples preferentially either to an aligned or anti-aligned spin.~\cite{meier2012optical} Rashba and Dresselhaus spin orbit Hamiltonians in principle serve as two internal magnetic fields~\cite{ganichev2004experimental,chang2005effect} that orient the electron spin polarization vector to influence the overall light absorption. To manifestly demonstrate the combined effect of Rashba and Dresselhaus Hamiltonians, we show in Fig.~\ref{rdallp} that the circular dichroism (CD) is reduced compared to the case (Fig.~\ref{nodress}) where there was no competing Dresselhaus effect. We also observe that the value of CD around the $ k = 0 $ point is no longer close to unity which suggests that light of both right and left handedness can be absorbed. Further, when $ \kappa = \alpha_{D}/\alpha_{R} $ is increased, implying a stronger contribution of the Dresselhaus Hamiltonian, circular dichroism drops. 
\begin{figure}[t]
\includegraphics[scale=0.8]{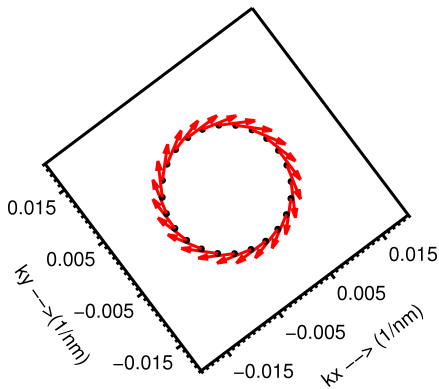} 
\caption{The clockwise spin distribution for a topological surface state present in a 20.0 $ \mathrm{nm} $ thick Bi$_{2}$Se$_{3}$ film. The spin is locked to momentum shown by the tangential lines on the plot. The surface state is at 0.1 $ \mathrm{eV} $ and energetically above the Dirac point.} 
\label{spinp}
\end{figure}
It would therefore be interesting to see if there exists values for $ \kappa $ such that circular dichroism drops below zero to acquire a negative value. Before we examine such a possibility and its physical implications, it is worthwhile to note that surface states (in momentum-space) in a topological insulator above and below the Dirac point possess exactly opposite helicity.~\cite{mciver2012control} Figure~\ref{spinp} shows the clockwise distribution of the spin polarization vector locked perpendicular to the momentum vector for a topological surface state of 20.0 $ \mathrm{nm} $ thick Bi$_{2}$Se$_{3}$ film. The surface state is located at 0.1 $ \mathrm{eV}$, placing it above the Dirac cone at 0.029 $ \mathrm{eV} $. For this arrangement, the circular dichroism at a $ k $-point in vicinity of $ \Gamma $ and above the Dirac cone would lie in the range $ \eta \in \left(0,1\right) $ while its exact negative energy counterpart below would exhibit an identical value with reversed sign. To relate to the TI model presented in this letter, what this means is that if we are able to show dual values for circular dichroism for an energy state above the Dirac point, the occurrence of which is precluded in the ``normal" Bi$_{2}$Se$_{3}$ TI film, a smooth transition in preferential absorption of left and right circularly polarized light would have been accomplished. This double-valued circular dichroism is explicitly shown in Fig.~\ref{rdpn}.   

\begin{figure}
\includegraphics[scale=0.75]{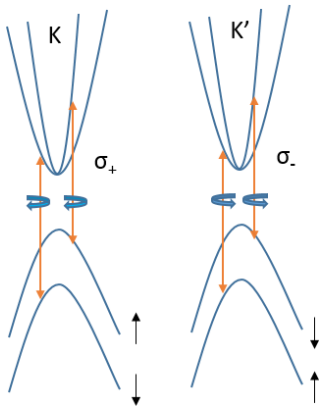} 
\caption{The time reversal symmetric K and K$^{'}$ valleys of MoS$_{2}$. Selective carrier excitations occur at K and K$^{'}$ valley edges. The bands are spin reversed and exactly at the edge absorb light of opposite handedness $\left(\sigma_{+}/\sigma_{-}\right) $ exclusively.} 
\label{timeval}
\end{figure}

To draw a chief conclusion from Fig.~\ref{rdpn}, firstly, we note that the single Dirac cone in TIs such as  Bi$_{2}$Se$_{3}$ and  Bi$_{2}$Te$_{3}$ does not allow one to achieve a dual circular dichroism unlike two-dimensional graphene or transition metal dichalcogenides(TMDCs) that have valley edges at equal energy but linked via time reversal symmetry. The schematic in Fig.~\ref{timeval} illustrates this point. The two valley edges K and K$^{'}$ follow an optical selection rule~\cite{xiao2012coupled,mak2012control} to absorb light of exactly opposite polarization. The Hamiltonian in Eq.~\ref{ham1} that describes the TI surface states, while do not exhibit such exclusivity at the $ \Gamma $ point, nonetheless, allows light of opposite polarization to be absorbed in varying quantities. Hence, the spin-dependent absorption pattern, as shown in Fig.~\ref{rdpn}, follows directly from the mutual strength of the Rashba and Dresselhaus coupling parameters which dictates the resultant in-plane spin orientation.
\begin{figure}
\includegraphics[scale=0.8]{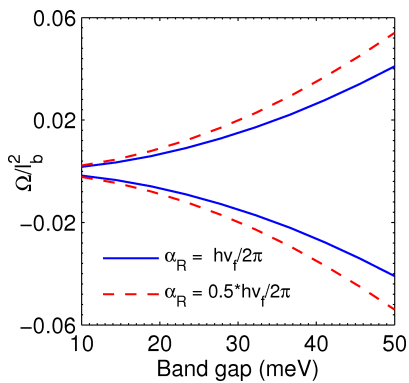} 
\caption{The two chiral components of Berry curvature with the Dresselhaus component in Eq.~\ref{berrycurv} set to zero. The Fermi velocity is 5 $ \times 10^{5} $ m/s and the $ k $-vector is set to $ 10^{-3} $ 1/\AA. The Rashba coefficient is indicated on the plot. Note that the Berry curvature on the ordinate axis is normalized to square of the magnetic length $ l_{b} = \sqrt{\dfrac{\hbar\,c}{eB}}$. } 
\label{bcurvf}
\end{figure} 
Another possible way to look at this is from the perspective of the emerging field of valleytronics~\cite{nebel2013valleytronics} in TMDCs that seeks to isolate electrons in the two distinct K and K$^{'}$ valleys based on the oppositely directed anomalous velocity~\cite{xiao2010berry} that arises on account of a finite Berry curvature. The Berry curvature, when time reversal symmetry is preserved,  has reversed signs in the two valleys. In our case, while we have a single copy of the Dirac cone, the surface electrons can be still be fictitiously described as belonging to two groups similar to those in K and K$^{'}$ valleys characterized by their opposite circular dichroism or Berry curvature. In fact, circular dichroism can be connected to Berry curvature as~\cite{xu2014spin}
\begin{equation}
\eta\left(k\right) = -\dfrac{\Omega\left(k\right)\,\cdot \hat{z}}{\mu^{*}_{B}\left(k\right)}\dfrac{e}{2\hbar}\Delta\left( k \right), 
\label{dicberr}
\end{equation} 
where $ \Omega\left(k\right) $ is the Berry curvature and $ \mu^{*}_{B} = e\,\hbar/2m^{*} $ is the spin Bohr magneton. The Berry curvature, following the procedure outlined in Ref~\onlinecite{sengupta2015influence} for the Hamiltonian of Eq.~\ref{ham1} and the related wave functions is
\begin{equation}
\Omega\left(k\right) = \mp\,\chi\dfrac{\left(\alpha^{2}_{R}-\alpha^{2}_{D}\right)\left[\alpha_{D}^{2} + \alpha_{R}^{2} + 2\alpha_{D}\alpha_{R}sin2\theta\right] }{\left(\alpha_{D}cos\theta + \alpha_{R}sin\theta\right)^{2} + \left(\alpha_{R}cos\theta + \alpha_{D}sin\theta\right)^{2}},
\label{berrycurv}
\end{equation} 
where $ \chi = \dfrac{\Delta}{2\left(\Delta^{2}+\vert \beta \vert^{2}\right)^{3/2}}$ and the upper(lower) sign for the Berry curvature corresponds to the eigen states $ E_{+}\left(E_{-}\right) $ of the model Hamiltonian in Eq.~\ref{ham1}. The magnetic length normalized Berry curvature where we have ignored the Dresselhaus contribution $\left(  \alpha_{D} = 0 \right) $ to avoid complications arising out of band crossing and degeneracy is shown in Fig.~\ref{bcurvf}. Also, note that the Berry curvature expression in Eq.~\ref{berrycurv} reduces to the familiar form~\cite{sengupta2015evaluation} when $ \alpha_{D} = 0 $.  

We can therefore infer from Eq.~\ref{dicberr} that the chiral Berry curvature switches sign (in an opposite sense) with circular dichroism when modulated through the spin-orbit Hamiltonian parameters. This switching, revealed as a toggling between the absorption of circularly polarized light of flipped polarization mimics the optical selection rule operational at the valley edges of TMDCs.

To conclude, we have shown that through a choice of the Rashba and Dresselhaus parameters, the inherent chiral nature of the surface states can be controlled to give rise to spin-assisted vertical optical transition observed via circular dichroism. 

This work was supported in part by the BU Photonics Center and U.S. Army Research Laboratory through the Collaborative Research Alliance(CRA).

\end{document}